\documentclass[12pt]{article}
\usepackage{times}
\usepackage{geometry,graphicx}
\usepackage[utf8]{inputenc}
\usepackage{enumitem,amssymb,wasysym}
\usepackage{ragged2e}
\newlist{thematic}{itemize}{8}
\setlist[thematic]{label=$\square$}
\usepackage{pifont}

\usepackage{natbib} 
\usepackage{paralist}
\let\olditem\item                                     
\renewenvironment{thebibliography}[1]{%
  \section*{\refname}
  \let\par\relax
  \renewcommand{\item}[1][]{\olditem[\textbullet]}%
\inparaenum}{\endinparaenum}

\begin{document}
\raggedright
\huge
Astro2020 Science White Paper \linebreak

Assembly of the Most Massive Clusters at Cosmic Noon \linebreak
\normalsize

\noindent \textbf{Thematic Areas:} \hspace*{60pt} $\square$ Planetary Systems \hspace*{10pt} $\square$ Star and Planet Formation \hspace*{20pt}\linebreak
$\square$ Formation and Evolution of Compact Objects \hspace*{31pt} $\CheckedBox$ Cosmology and Fundamental Physics \linebreak
  $\square$  Stars and Stellar Evolution \hspace*{1pt} $\square$ Resolved Stellar Populations and their Environments \hspace*{40pt} \linebreak
  $\CheckedBox$    Galaxy Evolution   \hspace*{45pt} $\square$             Multi-Messenger Astronomy and Astrophysics \hspace*{65pt} \linebreak
  
\textbf{Principal Author:}

Name:	Jeyhan S. Kartaltepe
 \linebreak						
Institution:  Rochester Institute of Technology
 \linebreak
Email: jeyhan@astro.rit.edu
 \linebreak
Phone:  585-475-7514
 \linebreak
 
\textbf{Co-authors:} (names and institutions)
  \linebreak
Caitlin Casey (UT Austin), Mark Dickinson (NOAO), Nimish Hathi (STScI), Anton Koekemoer (STScI), Brian Lemaux (UC Davis), Marc Postman (STScI), Gregory Rudnick (Univ. of Kansas) \\

\vspace*{0.2in}
\textbf{Abstract  (optional):}

Galaxy evolution is driven by many complex interrelated processes as galaxies accrete gas, form new stars, grow their stellar masses and central black holes, and subsequently quench. The processes that drive these transformations is poorly understood, but it is clear that the local environment on multiple scales plays a significant role. Today's massive clusters are dominated by spheroidal galaxies with low levels of star formation while those in the field are mostly still actively forming their stars. In order to understand the physical processes that drive both the mass build up in galaxies and the quenching of star formation, we need to investigate galaxies and their surrounding gas within and around the precursors of today's massive galaxy clusters -- protoclusters at $z\gtrsim2$. The transition period before protoclusters began to quench and become the massive clusters we observe today is a crucial time to investigate their properties and the mechanisms driving their evolution. However, until now, progress characterizing the galaxies within protoclusters has been slow, due the difficulty of obtaining highly complete spectroscopic observations of faint galaxies at $z\gtrsim2$ over large areas of the sky. The next decade will see a transformational shift in our understanding of protoclusters as deep spectroscopy over wide fields of view will be possible in conjunction with high resolution deep imaging in the optical and near-infrared.

\pagebreak

\justify
Galaxy evolution is driven by many complex interrelated processes. Galaxies accrete gas from the intergalactic/circumgalactic medium (IGM/CGM), form new stars from their gaseous interstellar media (ISM), and at the end of their lives, these stars return gas enriched with metals to the ISM/CGM. The processes that drive the triggering and subsequent quenching of galaxies is poorly understood, but it is clear that the local environment on multiple scales plays a significant role. In order to understand these physical processes that drive the quenching of star formation in the universe we need to investigate galaxies and their surrounding gas within and around the precursors to today's massive galaxy clusters -- protoclusters at $z\gtrsim2$.

\vspace*{-0.2in}
\section{Large Scale Structure and the Role of Protoclusters} 
\vspace*{-0.1in}

A critical outstanding problem in galaxy evolution is understanding environment's role on the formation and growth of galaxies in overdense and underdense regions of large scale structure. In the local universe ($z\sim0$), the final result of galaxy processing is widely observed in dense galaxy cluster environments. Ample evidence supports that galaxies living in cluster environments have more spheroidal-dominated morphologies, older stellar populations, and lower star formation rates (SFRs) relative to mass-matched galaxies in the field (\citealt{balogh98a,wake05a,skibba09a,collins09a}). While the SFRs of all galaxies are increasing towards higher redshifts  \citep{mada14}, a relation between redder colors (or lower SFRs) in cluster galaxies exists all the way to $z\sim 1.5$ (\citealt{mcoopz10, rudnick12, kovac14, lem18}).  This in turn indicates that a substantial amount of stellar mass buildup, and the subsequent quenching of star formation must have happened at $z>1.5$.  Indeed, there is tantalizing evidence that the cores of some $z>1.5$ clusters are experiencing elevated rates of star formation with respect to the field (e.g., \citealt{tran10, brodwin13, santos14, santos15}), although this has yet to be verified through systematic searches of representative samples of overdensities probing the galaxy populations that inhabit them.

This observational evidence is corroborated by results from simulations which indicate that forming clusters, i.e., protoclusters, form the vast majority of their stellar content at early times, $\sim50\%$ during the 1.5 Gyr period from $2<z<4$ \citep{chiang17}. During these epochs, it is predicted that protocluster environments become an important contribution to the overall comoving cosmic SFR density (SFRD), which is seen to peak at these redshifts. While protoclusters fill only $\sim3\%$ of the comoving volume of the universe at these epochs, they are estimated to contribute $20-30\%$ to the overall SFRD \citep{chiang17}, implying a rate of stellar mass assembly far outpacing that of the field. Other simulations of cosmological structure also show that galaxies grow faster in dense environments, eventually coalescing into massive galaxy clusters at $z<1$ \citep{collins09a,moster13a}. Quantifying observational trends such as the SFR or color density relations as a function of environment, redshift, and stellar mass, combined with the knowledge of density fluctuations in the very early Universe, have formed the backbone of our understanding of hierarchical growth and galaxy formation \citep{springel05a} but have largely been confined to studies at $z<1.5$ with the exception of a handful of studies on specific structures  \citep[e.g.,][]{steidel98a,venemans07a,doherty10a, cucciati18}. {\em The coming decade is the time to undertake a systematic observational mapping of overdensities at  $z\gtrsim2$ allowing us to empirically link early Universe density fluctuations to $z\sim0$ clusters and voids, something that is only now possible thanks to the promised imaging and spectroscopic survey power over wide areas provided by facilities such as Euclid, WFIRST, ATLAS, and the ELTs.}

\begin{figure}[t]
\vspace*{-0.4in}
\hspace*{-0.2in}
\begin{minipage}{0.55\textwidth}
    \includegraphics[width=3.5in]{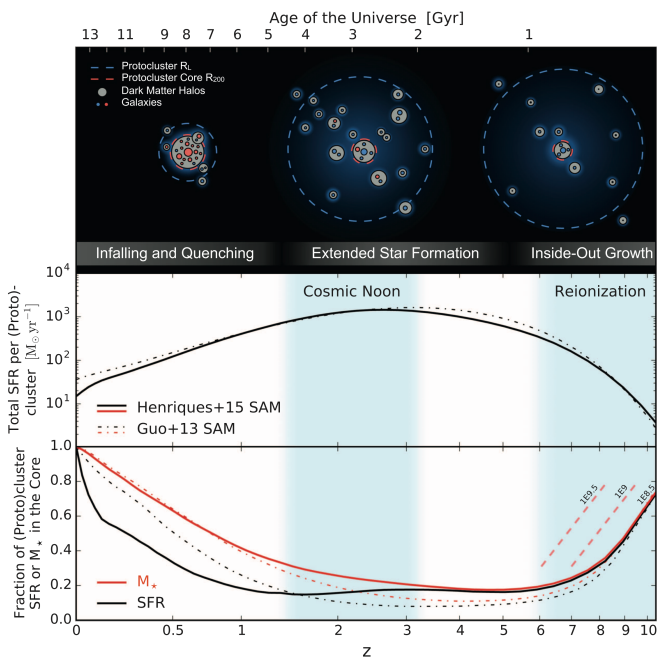}
\end{minipage}
\begin{minipage}{0.45\textwidth}
\caption{\small \textbf{Adapted from Chiang et al.~2017} \textit{Top Panel -- } The distribution of galaxies around protoclusters in various stages of cluster growth as determined from a semi-analytic model.  \textit{Middle Panel --} The total amount of star formation in the cluster.  \textit{Bottom Panel --} The fraction of the star formation that is in the protocluster core. At $z>1.5$ protoclusters are at the peak of their star formation histories (SFH) but that star formation is very extended.  Many galaxies and infalling halos are at both large physical radii and far outside of the virial radius of the protocluster core.  The SFR is correspondingly widely distributed. Tracking the growth of galaxies in protoclusters requires wide-field spectroscopic studies.  Studying extended protoclusters requires going to low masses, where environmental quenching is thought to dominate, and to high enough number densities (faint magnitudes) to measure galaxy overdensities -- requiring surveys over wide fields of view. }
\end{minipage}
    \label{Chiang}
     \vspace*{-0.3in}
\end{figure}

\vspace*{-0.6in}
\section{Identifying and Mapping Structures}
\vspace*{-0.1in}

At $z\lesssim 1.5$, and in a few exceptional cases at higher redshift \citep{gobat11, taowang16}, overdensities of galaxies are possible to unambiguously identify through their hot diffuse medium, which is manifest either through X-ray emission \citep[see review of ][]{kratsov12a}, or in the sub-mm through the Sunyaev-Zel'dovich (SZ) effect \citep[e.g.][]{menanteau09a}. Optical/near-infrared (NIR) searches of galaxy overdensities at these low redshifts are generally aided by the higher projected densities of galaxies within these overdense environments. With the cores of the structures having already collapsed to sizes with $r_{\rm ef}=2\,$Mpc.

But at earlier times, when processes generating the IGM have had less time to act and the structures have not had sufficient time to collapse and mature, observational signatures in the X-ray or through the SZ effect are generally not available. Furthermore, overdensities at $z>2$ occupy volumes many times larger than at $z\sim0$ \citep{onorbe14a}, i.e., typical progenitors of galaxy clusters are predicted to {\it subtend between a sixth and a third of a degree on the sky}. To attempt to circumvent these issues, signposts such as quasars \citep[e.g.,][]{miley04} or other AGN and dusty star forming galaxies (DSFGs) \citep[e.g.,][]{casey15a} are used to identify protoclusters. While these techniques are promising and can be efficient in certain cases, it is not clear that such signposts trace out typical protocluster environments rather than those in a special phase of their evolution. Rather, in order to make appreciable progress on open questions related to the formation and evolution of protoclusters and their galaxy populations, techniques which are able to select unbiased samples of both are required.

\vspace*{-0.2in}
\subsection{Galaxy Density Mapping}
\vspace*{-0.1in}

We can use galaxies themselves as tracers of protocluster environments and the surrounding density field through a combination of optical/NIR imaging and spectroscopy, and then translate back to the underlying matter density field through knowledge of the bias parameter of galaxy populations used in the density mapping. Such a technique is currently limited at $z\gtrsim2$ to tracing out the density field of star forming galaxies. As star forming galaxies are the dominant galaxy population at these epochs at essentially all stellar masses \citep[e.g.,]{ilbert13, tomczak14} a census of this population allows for a nearly unbiased tracing of the matter density field. The beginnings of systematic searches over relatively large areas of the sky (e.g., COSMOS) are starting to provide promising samples of protoclusters \citep[e.g.,][]{cucciati18}. Detections using such methods are, however, complicated by the presence of background and foreground objects that can quickly overpower the density peaks at these redshifts if extreme care is not taken. 

The deep broadband optical/NIR imaging that will be provided by {\it Euclid} \citep{euclidrb} and {\it WFIRST} \citep{wfirst19}, along with the highly multiplexed, large field of view optical/NIR multi-object spectroscopy possible on the ELTs, will deliver large samples of well-characterized protoclusters over the next decade by combining dense and highly complete spectroscopic coverage along with high quality photometric redshifts over large areas of the sky.

\vspace*{-0.2in}
\subsection{IGM Tomography}
\vspace*{-0.1in}

Observationally, $\sim$80\% of the baryonic matter in the universe predicted to exist by $\Lambda$CDM is missing. Cosmological hydrodynamical simulations indicate that these missing baryons are coursing through the IGM in a vast filamentary network of diffuse gas connecting large-scale luminous structures. While a few detections of this tenuous plasma have been reported, observers have yet to confirm the theoretical predictions. ELTs will provide the first opportunity to not only detect a signal from the IGM, but measure its density, structure, and chemical composition. 

If we want to observationally constrain the growth of structure, and the physics driving the accelerated growth of galaxies inside overdensities, we need an unbiased mapping of large scale structure. This requires deep spectroscopic campaigns over large areas of the sky in the optical and near-infrared, which will be enabled for the first time by facilities such as the ELTs, PFS on Subaru, the Mauna Kea Spectroscopic Explorer, and space missions such as ATLAS.
Spectroscopic observations are needed to identify galaxies that sit within discrete structures and to provide the deep continuum spectroscopy needed to map out foreground absorption from the Ly$\alpha$ forest.  This tomographic mapping of the intergalactic medium \citep[IGM;][]{lee14a,dlee14b} identifies gas absorption features in and around the high redshift galaxies that will be mapped in emission.  \citet{lee14a} demonstrate that a background source density of 360\,deg$^{-2}$ to 24$^{th}$ magnitude in $g$-band can be used to reconstruct 3D maps of the IGM on 2.5\,Mpc scales, and the existing density of 26$^{th}$ magnitude sources can reconstruct density maps down to 100\,kpc scales. The latter will only be possible with the ELTs to reach sufficient SNR on the rest-frame UV continuum emission of galaxies at $z>3$.  So far, the CLAMATO survey on Keck \citep{lee16a} has demonstrated the technique by mapping the $z\sim2.5$ cosmic web over 0.8\,deg$^2$, but deeper observations over wider FOVs will be essential.

\begin{figure}
    \centering
\begin{minipage}{0.5\textwidth}
    \includegraphics[width=3in]{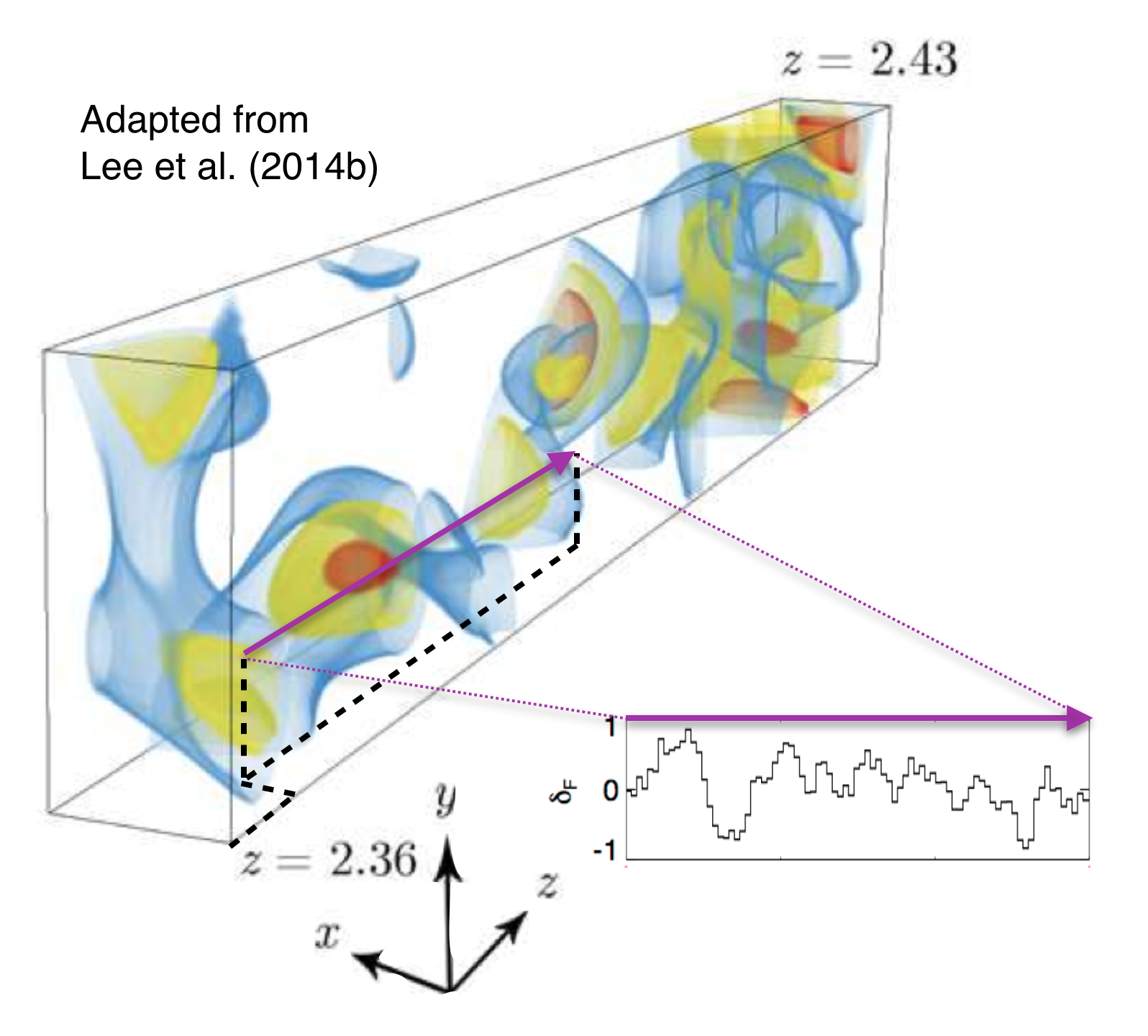}
    \vspace*{-0.2in}
\end{minipage}
\begin{minipage}{0.4\textwidth}
\hspace*{-0.1in}
\caption{
    A sample 3D map of the intergalactic medium (IGM) at $2.36<z<2.43$ in the COSMOS field as mapped by Ly$\alpha$ forest absorption in 24 background galaxies, pioneered in \citet{dlee14b}.  A sample skewer is highlighted with the purple arrow and cutout panel showing Ly$\alpha$ forest absorption along the line-of-sight.  Figure adapted from \citet{dlee14b}.}
    \label{clamato}
\end{minipage}
\end{figure}

Figure~\ref{clamato}, adapted from \citet{dlee14b}, shows an IGM Ly$\alpha$ forest absorption map from $2.36<z<2.43$ in the COSMOS field constructed with only 24 bright background emission sources surveyed in half a night of data with LRIS.  The structure seen at the edge of the box at $z\ge2.43$, described in \citet{lee16a}, has also been detected in emission by analysis of spectroscopic redshifts for known LBGs and DSFGs \citep{diener15a,chiang15a,casey15a}, and is estimated to grow to a 10$^{15}$\,M$_\odot$ structure by $z=0$.  While such protoclusters are now being identified regularly in extragalactic deep fields given the large volumes they probe across large redshift ranges \citep{chapman09a,yuan14a,hung16a,casey16a}, only this structure has been directly detected via absorption using IGM tomography.

Ultimately, we will be able to combine measurements of the LSS galaxy density maps from photometric surveys and IGM tomography to analyze the correlation of underlying dark matter structure -- to which IGM tomography is sensitive -- with large scale baryonic structure.  For example, some theories predict certain correlations between galaxy alignment and galaxy bias with the underlying dark matter structure that is so far unconstrained \citep{aragon-calvo07a,hahn07a,jones10a,codis12a,tempel13a,dubois14a}.  Higher-redshift studies of the cosmic web are particularly important as their components are not fully evolved and gravitationally merged, and much information regarding the properties of galaxies and dark matter halos, that are lost at low-$z$ due to the gravitational non-linear-interaction regime, is still intact at $z>2$ \citep{jones10a,cautun14a}. This sets the need for contiguous large-volume surveys at higher redshifts (in order to limit the cosmic variance per field), which need to be equipped with very accurate photometric redshifts.

\vspace*{-0.2in}
\section{Galaxy Properties}
\vspace*{-0.1in}

In order to understand the physical processes in  high redshift galaxies, it is imperative to investigate their spectroscopic properties.  Rest-frame UV spectroscopic observations are a great probe of Ly$\alpha$ emission, ISM/CGM properties, and galaxy outflows, while rest-frame optical observations are crucial for investigating star formation rates (SFRs), dust properties, and metallicities. The rest-frame UV luminosity function at $z\sim2$ has a steep faint-end slope \citep[e.g.,][]{alav16}, which implies that the galaxy population at $z\sim2$ is dominated by galaxies that are fainter than $\sim$~25--26 AB mag. It is very difficult to spectroscopically observe  these \emph{faint} galaxies using current 8--10m class telescopes; thus they are the best targets for deep observations with ELTs. Through extensive study of faint galaxies at the peak epoch of star formation activity \citep[e.g.,][]{mada14}, using thorough correlations between rest-frame UV/optical spectral features, SED-based physical parameters, and morphological signatures, we will be able to observationally distinguish between different models of galaxy assembly and growth, allowing us to bridge the gap between the very well-studied z$\sim$0 galaxies and higher redshift samples.

\begin{figure}[!]
\vspace*{-2.3in}
\hspace*{-0.2in}
\begin{minipage}{0.75\textwidth}
    \includegraphics[width=5in]{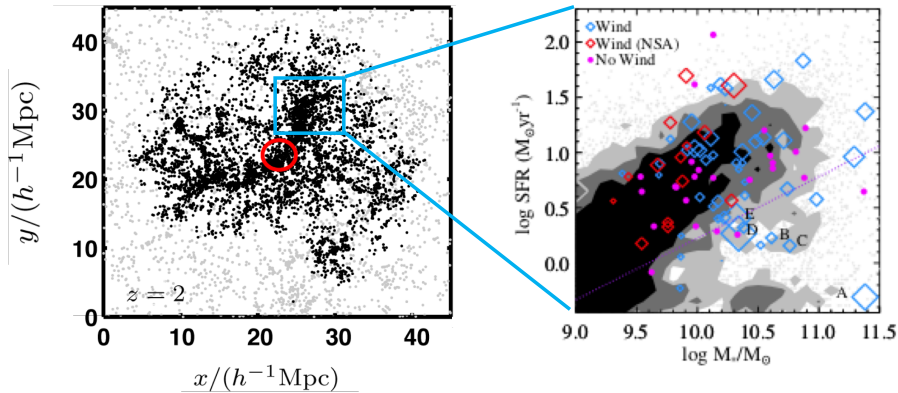}
\end{minipage}
\begin{minipage}{0.25\textwidth}
\caption{\small Adapted from Muldrew et al.~2015 (left) and Rubin et al.~2014 (right). Illustration of simulated large scale structure surrounding a massive protocluster at $z\sim 2$ and an example of the types of measurements that our program will make over the extendend filamentary structure.}
\end{minipage}
    \label{MuldrewRubin}
    \vspace*{-2.2in}
\end{figure}

\emph{Continuum and Absorption Line Spectroscopy}:
High S/N features in the rest-frame UV spectra of star forming galaxies provide deep understanding into the physical properties of their massive stars and gas, including the multi-phase ISM and CGM. The low- and high-ionization interstellar absorption lines provide better understanding into various IGM/CGM properties including, the gas covering fraction, which is directly linked to Ly$\alpha$ properties and escape of Lyman continuum ionizing radiation, and the kinematic signatures of galaxy outflows (see, for example, Fig.~3), which play a vital role in the formation and evolution of galaxies. 

At the same time, stellar continuum spectroscopy (e.g., Balmer absorption lines, the 4000\AA\ break) of post-starburst or quiescent galaxies, which typically have weak or no emission lines, is critically important for measuring stellar ages and constraining star formation histories at z$\gtrsim$2. Such studies require high S/N continuum spectroscopy, which has been only attempted for the brightest galaxies at z$\gtrsim$2 because it is very challenging for 8--10m class telescopes to reach that depth in a reasonable amount of time. The large collecting area combined with moderate-to-high resolution and multiplexed optical and NIR spectrographs on the ELTs is essential for such deep observations.

\emph{Emission Line Spectroscopy}:
The Ly$\alpha$ line is the strongest UV emission line in star forming galaxies and a very important spectroscopic signature to confirm redshift. Ly$\alpha$ photons are produced by recombination in H II regions and then propagate through the ISM, interacting with both neutral hydrogen and dust particles. Because of its resonant nature, Ly$\alpha$ photons can help us to understand H II regions as well as the much extended ISM/CGM through which they random-walk their way out of galaxies. The shape of the Ly$\alpha$ profile (e.g., number of peaks, asymmetry) can also be used to study the kinematics and density distribution of outflowing neutral gas \citep[e.g.,][]{verh06,kula12}. In addition to Ly$\alpha$, nebular emission lines including, C III] $\lambda$1909, He II $\lambda$1604, and [O III] $\lambda$1661/1666, are produced in H II regions and are useful for probing the ionized ISM and radiation field produced by massive stellar populations \citep[e.g.,][]{erb10, cass13, star14, lefe18, naka18}.

Rest-frame optical emission lines also play a significant role in our understanding of various physical properties of galaxies. With NIR multi-object spectrographs, we can probe [OII] $\lambda$3727, H$\beta$ $\lambda$4861, [OIII] $\lambda$5007, [NII] $\lambda$6548, and H$\alpha$ $\lambda$6564 for galaxies between $z\sim1.5$ to $z\sim5.3$ ($z\sim2.6$ to capture the full suite of lines). These emission lines will be used to estimate dust attenuation (H$\alpha$/H$\beta$), SFRs (H$\alpha$, H$\beta$, [OII]), and line diagnostics to detect intrinsically weak / obscured AGN and constrain physical ISM conditions, such as metallicity and ionization parameter. With ELTs, we will be able to make these measurements for fainter, lower-mass systems than is possible with 8--10 m class telescopes, enabling us to trace galaxy and ISM properties throughout protoclusters and their surrounding filaments for galaxies and structures that will evolve into those that we observe today in the nearby universe.

\pagebreak

\end{document}